\documentclass[aps,prd,twocolumn,nofootinbib,superscriptaddress,10pt,floatfix]{revtex4-1}
\usepackage{amstext}
\usepackage{amssymb}
\usepackage{amsmath}

\usepackage[utf8]{inputenc}
\usepackage{color}
\usepackage{graphicx}

\usepackage{array}
\usepackage{subfigure}
\usepackage{color}
\usepackage{listings}
\usepackage{natbib}
\usepackage{url}
\usepackage{bm}
\usepackage{multirow}
\usepackage{etoolbox}
\usepackage[utf8]{inputenc}
\usepackage[nolist]{acronym}
\usepackage[breaklinks, plainpages=false, colorlinks=true, anchorcolor=cyan, linkcolor=red, citecolor=cyan, urlcolor=magenta, bookmarks=false]{hyperref}
\usepackage[normalem]{ulem}
\usepackage{tabu}
\usepackage{tikz}
\usetikzlibrary{arrows,shapes,trees,decorations.pathreplacing}
\usepackage{import}
\usepackage{svn-multi}
\usepackage{lineno}
\usepackage{hyperref}
\hypersetup{
    colorlinks=true,
    linkcolor=blue,
    filecolor=magenta,
    urlcolor=cyan,
}

\newcommand{\chieff}{\ensuremath{\chi_{\mathrm{eff}}}}
\newcommand{\rankingstat}{\ensuremath{\Lambda_s}}

\newcommand{\msun}{\ensuremath{\mathrm{M}_{\odot}}}

\begin{document}
\title[]{An optimized PyCBC search for gravitational waves from intermediate-mass black hole mergers}
\author{Koustav Chandra}
\affiliation{Department of Physics, Indian Institute of Technology Bombay, Powai, Mumbai 400 076, India}
\email{koustav.chandra@iitb.ac.in}
\author{V.~Villa-Ortega}
\affiliation{IGFAE, Campus Sur, Universidade de Santiago de Compostela, 15782 Spain}
\author{T.~Dent}
\affiliation{IGFAE, Campus Sur, Universidade de Santiago de Compostela, 15782 Spain}
\author{C.~McIsaac}
\affiliation{University of Portsmouth, Portsmouth, PO1 3FX, United Kingdom}
\author{Archana~Pai}
\affiliation{Department of Physics, Indian Institute of Technology Bombay, Powai, Mumbai 400 076, India}
\author{I.~W.~Harry}
\affiliation{University of Portsmouth, Portsmouth, PO1 3FX, United Kingdom}
\author{G.~S.~Cabourn~Davies}
\affiliation{University of Portsmouth, Portsmouth, PO1 3FX, United Kingdom}
\author{K.~Soni}
\affiliation{Inter-University Centre for Astronomy and Astrophysics, Pune 411007, India}


\begin{abstract}
The detection of \acp{IMBH} i.e.\ those with mass $\sim 100$--$10^5$\msun, is an emerging goal of \ac{GW} astronomy with wide implications for cosmology and tests of strong-field gravity. Current PyCBC-based searches for compact binary mergers, which matched filter the detector data against a set of template waveforms, have so far detected or confirmed several \ac{GW} events. However, the sensitivity of these searches to signals arising from mergers of \ac{IMBH} binaries is not optimal. Here, we present a new optimised PyCBC-based search for such signals. Our search benefits from using a targeted template bank, stricter signal-noise discriminators and a lower matched-filter frequency cut-off. In particular, for a population of simulated signals with isotropically distributed spins, we improve the sensitive volume-time product over previous PyCBC-based searches, at an \acl{IFAR} of 100 years, by a factor of 1.5 to 3 depending on the total binary mass. We deploy this new search on Advanced LIGO-Virgo data from \acl{O3a}.  The search does not identify any new significant \ac{IMBH} binaries but does confirm the detection of the short-duration \ac{GW} signal GW190521 with a false alarm rate of 1 in 727 years.
\end{abstract}

\maketitle

\acrodef{GW}{gravitational-wave}
\acrodef{IMBH}{intermediate-mass black hole}
\acrodef{FAR}{false alarm rate}
\acrodef{IFAR}{inverse false alarm rate}
\acrodef{SNR}{signal-to-noise ratio}
\acrodef{O3a}{the first half of the third observing run}
\acrodef{PSD}{power spectral density}
\acrodef{GWTC-2}{second gravitational wave transient catalogue}
\acrodef{CBC}{compact binary coalescence}
\acrodef{BBH}{black hole binary}
\acrodef{GWTC}{gravitational wave transient catalogue}
\acrodef{BH}{black hole}

\section{Introduction}
\label{sec:intro}

What started as a trickle back in September 2015~\citep{Abbott:2016blz} has turned into a veritable deluge~\citep{LIGOScientific:2018mvr, Abbott:2020niy} of~\acf{GW} detections at the end of~\acf{O3a}  of advanced~\ac{GW} detectors. As a result, the gravitational wave catalogues~\citep{LIGOScientific:2018mvr, Nitz:2018imz, Nitz:2019hdf, Venumadhav:2019lyq, Venumadhav:2019tad, Abbott:2020niy, Nitz:2021uxj} now list more than $50$~\acfp{CBC}. This rapid surge in~\ac{GW} events is largely due to the upgrades of the Advanced LIGO~\citep{TheLIGOScientific:2014jea} and Advanced Virgo~\citep{TheVirgo:2014hva} instruments in the interim between the observing runs. The search algorithms, namely cWB~\citep{Klimenko:2004qh, Klimenko:2005xv,Klimenko:2006rh, Klimenko:2011hz, Klimenko:2015ypf}, GstLAL~\citep{ Messick:2016aqy, Sachdev:2019vvd, Hanna:2019ezx}, PyCBC~\citep{Usman:2015kfa, Nitz:2017svb, Davies:2020tsx}, MBTA~\citep{Adams:2015ulm, Aubin:2020goo} and SPIIR~\citep{chu2017low} also have improved significantly enabling them to better separate gravitational signal from the detector noise. 
As a consequence, the \acf{GWTC-2} not only reports the detection of a plethora of stellar mass compact binaries~\citep{Abbott:2020uma, Abbott:2020niy} but also details the detection of a couple of relatively asymmetric-mass compact binaries~\citep{LIGOScientific:2020stg, Abbott:2020khf}, and a signal, GW190521, from a putative \acf{IMBH} binary~\citep{Abbott:2020tfl, Abbott:2020mjq}. 

The signal GW190521 is quite different from other reported \ac{GW} events. The pre-merger phase of the signal was barely observable within the detectors' bandwidth, allowing for its multivalent interpretation~\citep{Abbott:2020tfl, Abbott:2020mjq, CalderonBustillo:2020odh, Gayathri:2020coq, Palmese:2020xmk, Liu:2020lmi, CalderonBustillo:2020srq, Nitz:2020mga}. If we assume that the signal is from a quasi-circular black hole binary, then the heavier component of the binary has a high probability of populating the pair-instability mass gap, and the remnant, with mass $\sim 142$\,\msun, is the first conclusively observed intermediate-mass black hole~\citep{Abbott:2020tfl, Abbott:2020mjq}.

The event was first identified on 21 May 2019 at 03:02:29 UTC by the low latency matched-filter search for compact binaries, PyCBC-LIVE~\citep{Nitz:2018rgo, DalCanton:2020vpm}, and the weakly modelled transient search, cWB, in its \ac{IMBH} configuration~\citep{Szczepanczyk:2020osv}. The two searches reported the event with an \acf{IFAR} of 8.34 years and 28.0 years, respectively. The event was also reported by two other low-latency searches, namely GstLAL and SPIIR, though with lower statistical significance.

Low latency searches are designed to generate rapid alerts to enable prompt follow-up of gravitational-wave triggers for potential electromagnetic or other counterpart signals. Hence, the statistical significance estimated by low-latency pipelines is limited due to the available amount of empirically generated background. Therefore, most search pipelines re-scrutinise the strain data with improved calibration, data quality vetoes, and signal-noise discriminators, which are generally unavailable during low-latency searches. Further, offline searches estimate the background over a larger amount of data, and can thus provide a better estimate of a candidate's statistical significance. In particular, when such an analysis was carried out at the time around GW190521, the search pipelines reported a maximum~\ac{IFAR} of 4900 years~\citep{Abbott:2020tfl, Szczepanczyk:2020osv}.

However, contrary to other searches, the offline PyCBC-based search, PyCBC-broad, identified the event with an~\ac{IFAR} of less than a year.  This is primarily due to the strong disagreement between the event and the best-matching template used in the search~\citep{GW190521:discovery_supplement}. This search also suffered from an inherent ``look-elsewhere effect'' 
as it tries to find \acp{CBC} signals over a broad parameter space from binary neutron stars up to \ac{IMBH} binaries. The other PyCBC-based search, namely PyCBC-BBH, did not detect the event due to its tight tuning to signals closely matching stellar-mass \ac{BBH} templates.
\begin{figure*}[ht]
  \centering
    \includegraphics[width=\textwidth]{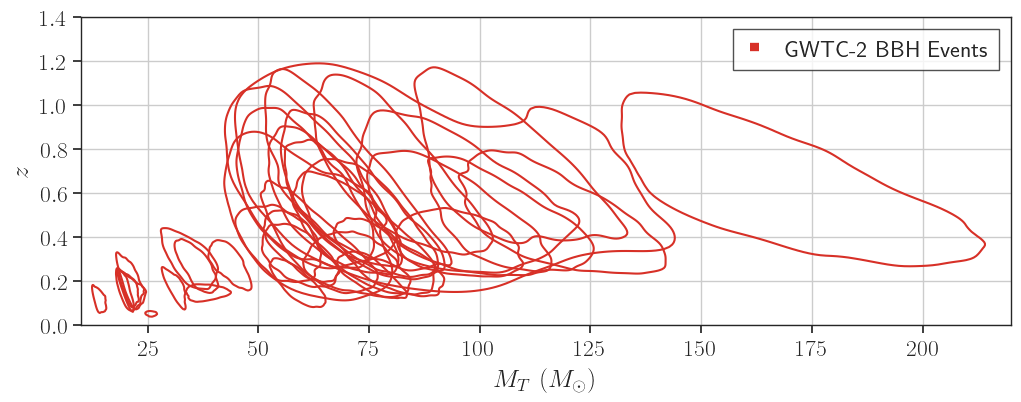}
\caption{Posterior probability distribution of the source frame total mass $M_T$ and redshift $z$ for the 37 confident \ac{BBH} events announced in \ac{GWTC-2} [red].  The contours enclose 90\% of the posterior probability, inferred under the default parameter estimation priors used in the \ac{GWTC-2} publications~\citep{Abbott:2020niy}.  For several high-mass mergers, the redshift extends to $z\sim 1$: note that the template masses required to detect such signals scale as $(1+z)M_T$.}
\label{fig:redshift}
\end{figure*}

The two gravitational-wave transient catalogues to date also show some interesting trends concerning binary masses.  In the first LIGO-Virgo catalogue, GWTC-1~\citep{LIGOScientific:2018mvr}, a lack of systems with $m_1 \gtrsim 45\,\msun$ is evident.  However, more recently \ac{GWTC-2} lists several systems in this range, indicating additional structure in the \ac{BBH} population distribution as discussed in \citet{Abbott:2020gyp}.  
In Fig.~\ref{fig:redshift} we show the inferred source total masses $M_T$ and redshifts $z$ for 37 confidently detected \acp{BBH} announced in \ac{GWTC-2}.  The lack of low-mass events at high $z$ is due to a selection effect as the amplitude of emitted \ac{GW} scales with mass. Already, several high-mass \acp{BBH} are detectable at $z \sim 1$: thus, the template masses required to detect such systems are significantly larger than the source masses due to redshifting of the \ac{GW} signal~\citep{Krolak:1987ofj}.  Thus, even for (relatively heavy) stellar-mass \ac{BBH} ($\gtrsim 50\,\msun$), an effective \ac{IMBH} binary search is required. 

We expect these trends to continue during upcoming runs when \ac{GW} interferometers achieve higher global network sensitivity to \ac{BBH} mergers~\citep{Abbott:2020qfu}. The improvement will facilitate the detection of potentially hundreds of \acp{BBH} per year, reaching out to larger redshifts~\citep{Mehta:2021fgz}. Also, within the turn of the next decade, we will have more sensitive detectors such as Cosmic Explorer~\citep{Reitze:2019iox} and the Einstein Telescope~\citep{Maggiore:2019uih, Sathyaprakash:2012jk} that will help detect more \ac{BBH} mergers at even further distances, possibly up to $z \sim 20$ and higher. The launch of the Laser Interferometer Space Antenna (LISA) will potentially allow us to look even deeper and for more massive \acp{BBH}~\citep{Audley:2017drz}. Such binaries may be heavier than those observed so far~\citep{Safarzadeh:2019ctv} and so would produce remnants in the lower~($10^2\,\msun$--$10^3\,\msun$) to mid~($10^3\,\msun$--$10^4\,\msun$) \ac{IMBH} mass range ~\citep{Mangiagli:2019sxg}. Detecting such systems in multiband \ac{GW} window~\citep{Sesana:2016ljz, Jani:2019ffg} will help us to estimate better the population distribution of low and mid mass range \acp{IMBH}, which is currently lacking. It will also enable us to understand better the formation mechanism of \ac{IMBH} systems and to perform more detailed tests of general relativity in the strong-field regime~\citep{Abbott:2020jks}. And as mentioned above, \ac{GW} signals from high redshift binaries will have higher detector frame masses, up to an order of magnitude or larger than the source masses. So, regardless of whether we detect any more \acp{IMBH}, we still need to extend the target space of current \ac{CBC} searches and tune them to prevent missing these potential signals.

We make such an attempt in this work. Here, we propose a new PyCBC-based search which is optimised to detect both \ac{GW} signals from \ac{IMBH} binaries and highly redshifted heavy stellar-mass \ac{BBH} mergers. We construct a bank of non-precessing waveforms simulating \ac{BBH} mergers with component detector frame masses $m_{1,2}(1+z) \geq 40\,\msun$, total detector frame masses $M_T(1+z) \in [100, 600]\,\msun$ and mass-ratios $q \in [1, 10]$. Because of our choices in the target search space, our search is susceptible to a larger number of noise triggers, requiring us to impose stricter noise vetoes and discriminators~\citep{Cabero:2019orq, Davis:2020nyf}. We find that, while keeping the basic structure of the PyCBC search algorithm, our modifications increase the search sensitivity, in terms of sensitive volume-time, for a simulated population of binaries, by an average factor of $\sim 2$ at an \ac{IFAR} of 100 years, compared to the PyCBC-broad search used in \citet{Abbott:2020tfl}.

We organise our paper as follows. Sec.~\ref{sec:review} serves as an overview of the multi-detector PyCBC analysis. Optimisations and changes to this analysis are presented in Sec.~\ref{sec:new-methods}. In Sec.~\ref{sec:evaluation}, we describe how these modifications combine to improve the sensitivity of the PyCBC analysis to \ac{IMBH} mergers. Because of the changes made, the search now recovers the heavier mass black hole binary mergers from the third observing run with higher significance, and we have summarised these findings in Sec.~\ref{sec:results}. Finally, in Sec.~\ref{sec:conclusion} we conclude the paper by indicating the future optimisation prospects in the upcoming observing runs.

\section{Review of Existing PyCBC detection methods for CBC systems}
\label{sec:review}
The strain that an incident \ac{GW} signal produces in the interferometric detector is a weighted linear combination of the gravitational wave polarisation $h_+(t)$ and $h_\times(t)$:
\begin{equation}
    s(t) = F_+(\alpha, \delta, \psi)h_+(t;\vec{\zeta},\vec{\kappa}) +F_\times(\alpha, \delta, \psi)h_\times(t;\vec{\zeta},\vec{\kappa})~,
\end{equation}
where the weights $F_{+/\times}$ represent the detector beam pattern functions. These functions depend on the source's sky location $(\alpha,\delta)$ with respect to the detector and on the polarisation angle $\psi$. If the signal is from a quasi-circular binary, then $\vec{\zeta}$ in the above equation describes the intrinsic parameters of the binary, which comprises of the individual component masses $m_i$ and the dimensionless spin $\vec{\chi}_i$ of the binary.  $\vec{\kappa}$, on the other hand, consists of the luminosity distance $D_L$ to the source, the inclination angle $\iota$ of the binary with respect to the observer and the merger time and phase $(t_c,\varphi_c)$ of the signal.

Current PyCBC-based \ac{GW} searches primarily target signals coming from mergers of binaries consisting of a pair of neutron stars, a pair of black holes, or a neutron star-black hole pair. So, in the subsequent subsections, we briefly review the steps involved in detecting these well-modelled \ac{CBC} signals.

\subsection{Pre-processing of detector data}
\label{subsec:preprocessing}
The calibrated detector output is an evenly spaced time-series data containing information about strain due to different sources. This makes it amenable to standard techniques of time series analysis. The key foundation of prevalent gravitational wave time-series analysis is a linear model of the data $d(t)$: $d(t) = s(t) + g(t) + n_G(t)$, where $s(t)$ is the \ac{GW} signal strain, $n_G(t)$ is the strain due to random (Gaussian) detector noise and $g(t)$ is the strain due to non-Gaussian terrestrial transients, a.k.a.\ `glitches'.

For most of the time, the data $d(t)$ is just $n_G(t)$ which is usually assumed to be additive, wide-sense stationary coloured Gaussian noise. This presupposition allows one to fully describe the noise's statistical properties in terms of its \ac{PSD}, $S_n(f)$. PyCBC-based searches estimate the \ac{PSD} via Welch's method and use it to whiten the detector data so that any excess power due to the presence of either $s(t)$ or $g(t)$ is more prominent.

However, the \ac{PSD} estimate can be significantly affected during periods of poor detector performance. Therefore, the PyCBC toolkit~\citep{alex_nitz_2021_4556907} uses data quality information to zero out flagged times of badly contaminated data before evaluating \ac{PSD}. Further, to reduce the number of false alarms caused due to high-amplitude glitches, the search algorithm internally identifies times of large deviation from Gaussian noise in the whitened strain data and window them out using a procedure known as auto-gating~\citep{Usman:2015kfa}. The PyCBC algorithm also vetoes out flagged times of additional glitches that can generate loud triggers but do not contaminate the surrounding data to affect the performance of the PSD estimation. This data wrangling process ensures better search performance but cannot completely purge the data from all glitches $g(t)$.

\subsection{Matched Filtering}
\label{subsec:MF}

Most current \ac{CBC} searches assume that the incident \ac{GW} signal can be adequately represented by the dominant harmonic of a quasi-circular non-precessing binary. The form of such a signal is well-modelled in the framework of general relativity. So, after the pre-processing step, PyCBC-based searches use matched filtering~\citep{Sathyaprakash:1991mt} to find whether or not there is a \ac{GW} signal buried in the detector data~\citep{Allen:2005fk, Babak:2012zx, Usman:2015kfa}. As the parameters of the incident signal are not known a priori, \ac{CBC} searches construct and search over a discrete \textit{bank} of templates spanning the search space. 

The detector response due to the dominant harmonic of a non-precessing quasi-circular binary in the frequency domain can be written as:
\begin{equation}\label{eq:mod-strain}
    \tilde{h} (f;\vec{\theta}) =\frac{F}{D_L} \tilde{h}_{22}(f;\vec{\zeta}) e^{i\phi}
\end{equation}
provided stationary phase approximation holds~\citep{Sathyaprakash:1991mt, Finn:1992xs}. Here, $\tilde{h}_{22}$ is the geocentric waveform, parameterised completely by $\vec{\zeta}$, which forms the set of filter waveforms used in the search. $F$, on the other hand, describes the effective antenna pattern function:
\begin{equation}
F = \sqrt{F_+^2\Bigg( \frac{1 + \cos^2 \iota}{2} \Bigg) + F_\times^{2}\cos^2 \iota}
\end{equation}
while 
\begin{equation}
    \phi = \arctan{\Bigg( \frac{2F_\times \cos \iota}{F_+\big(1 + \cos^2{\iota\big)}}\Bigg)} + \varphi_c + 2\pi f\Delta t
\end{equation}
represents the effective merger phase, with $\Delta t$ denoting the arrival time of the signal to the detector
relative to a geocentric time reference. 

The above presupposition also allow us to decompose the template strain, $\tilde{h}_{22}$, into two orthonormal components, $\hat{h}_c$ and $\hat{h}_s$, with an overall effective amplitude. These can be used to define the log likelihood ratio, $\ln \mathcal{L}$, between the signal and noise hypothesis given $g(t)=0$. It also allows one to maximise the log likelihood ratio over the effective template phase and amplitude, giving~\citep{Finn:1992wt}:
\begin{equation}\label{eq:max-likelihood}
    \ln \mathcal{L}_\textrm{max}(t; \vec{\zeta}) = \frac{1}{2} \rho^2(t; \vec{\zeta})\,.
\end{equation}
Here, $\rho(t)$ denotes the matched-filter \ac{SNR}:
\begin{equation}\label{eq:SNR}
    \rho(t; \vec{\zeta}) = \sqrt{\langle d | \hat{h}_s \rangle^2 + \langle d | \hat{h}_c \rangle^2},
\end{equation}
which is evaluated for each template independently in each detector. $\langle d | h \rangle$ defines the noise-weighted inner product between the pre-processed data and the template in the frequency domain and is given as:
\begin{equation}\label{eq:innerproduct}
    \langle d | h \rangle (t) = 4 \mathbb{R}e  \int_{f_\textrm{low}}^{f_\textrm{high}}\frac{\tilde{d}(f)^\ast \tilde{h}(f)}{S_n(f)}e^{2i\pi ft}\,df\,.
\end{equation}
Here, $f_\textrm{low}$ and $f_\textrm{high}$ represent the lower and upper frequency cut-off of the matched-filter operation, respectively. Typically, $f_\textrm{low}$ is set to the frequency from which the filter waveform is generated, while $f_\textrm{high}$ is set either as the final frequency cut-off of the template or the Nyquist frequency of the detector data, whichever is smaller. A peak in this \ac{SNR} time-series above a specific threshold will correspond to an astrophysical trigger, provided there is no glitch in the data. Since we do not know a priori the parameters describing the \acl{GW} in the data, current \ac{CBC} searches carry out this matched filter operation over all the templates in the bank. The template which best describes any signal present in the data gives the largest peak in the \ac{SNR} time-series: hereafter, we will call such a filter waveform the best-matched template. 

\subsection{Single Detector Noise Discriminators}
\label{subsec:single-triggers}

Matched filtering gives the optimal detection statistic provided there were no glitches in the data~\citep{Wainstein:1970}. Unfortunately, this is not the case. Pre-processing of the calibrated detector data eliminates loud glitches, but the relatively softer glitches still pollute the detector data and form the majority of the single-detector triggers. To reduce the effect of these glitches on the search, the PyCBC toolkit uses several noise discriminators at the single detector level.

Firstly, the PyCBC toolkit discards trigger below some pre-determined \ac{SNR} threshold and clusters triggers within a fixed window of time --- keeping only the triggers with the largest \ac{SNR}. It also tries to suppress loud glitches $g(t)$ by performing two different signal consistency tests: $\chi_\mathrm{r}^2$ \citep{Allen:2004gu} and $\chi_{\mathrm{sg}}^2$~\citep{Nitz:2017lco}. The $\chi_\textrm{r}^2$ test checks if the morphology of the candidate trigger in frequency domain is consistent with that of the best-matched template while the $\chi_{\textrm{sg}}^2$ test is designed to down-rank triggers with excess power at frequencies exceeding those allowed by the best-matched template. In the presence of Gaussian noise, both $\chi_\mathrm{r}^2$ and $\chi_{\mathrm{sg}}^2$ follow a normalised $\chi^2$ distribution.  However, these statistics produce large values in the presence of glitches or when the signal at hand is not well-modelled by the best-matched template. PyCBC-based searches use these test results to re-weigh the trigger \ac{SNR}~\citep{Babak:2012zx, Nitz:2017lco}. They first calculate the ``reweighted SNR'' for the triggers via
\begin{equation}
    \tilde{\rho} =
    \begin{cases}
        \rho \Big[\frac{1}{2} \big(1+(\chi^2_\textrm{r})^{3}\big)\Big]^{-1/6} & \mathrm{for\,} \chi^2_\textrm{r} > 1 \\
        \rho & \mathrm{for\,} \chi^2_\textrm{r} \leq 1
\end{cases}
\label{eq:newsnr}
\end{equation}
and then further re-weigh it by calculating
\begin{equation}
    \hat{\rho} = 
    \begin{cases}
        \tilde{\rho}(\chi^2_\textrm{sg})^{-1/2} & \text{if}~\chi_\textrm{sg}^2 > 4, \\
        \tilde{\rho} & \text{otherwise.}
\end{cases}
\label{eq:newsnr_sgveto}
\end{equation}
The above two amendments to the standard trigger \ac{SNR} help reduce the apparent loudness of the glitches and hence increase the \ac{FAR} of quieter short duration glitches. 

The detector data also contains broadband non-stationary detector noise, which affects the detector's sensitivity in the order of tens of seconds. To alleviate the effects of this noise, the PyCBC algorithm re-scales $\hat{\rho}$ with an estimate of the short-term variation in the \ac{PSD}, v$_s(t)$,~\citep{Mozzon:2020gwa}. When combined with $\hat{\rho}$, this gives the single detector statistic of the trigger:
\begin{equation}
    \Breve{\rho} = \hat{\rho}\textrm{v}_s(t)^{-0.5}\,.
    \label{eq:newsnr_sgveto_psdvar}
\end{equation}

\subsection{Multi-detector Coincident Candidates}
\label{subsec:coincidence}
After generating single-detector triggers from each detector, templated searches perform a coincidence test to identify candidate events. This involves checking whether triggers from the same template are observed across multiple detectors with time-delays consistent with the light travel time between detectors. 

Under the assumption that the noise is independent between detectors, this coincidence test helps us to further separate signal triggers from the noise background. We perform this test for all possible two or three detector combinations. To determine the significance of the surviving multi-detector candidates, PyCBC searches assigns them a ranking statistic based on the relative rate densities of signal and noise events~\citep{Nitz:2017svb,Davies:2020tsx}:
\begin{multline}
    \rankingstat = - \ln A_{\{a\}} - \sum_{a} \ln r_{a,i}(\Breve{\rho}_a) + \ln \frac{p(\vec{\Omega}|S)}{p(\vec{\Omega}|N)} 
    + 3\ln \frac{\sigma_\mathrm{min}}{\sigma_\mathrm{ref}}\,.
\end{multline}
The first two terms on the RHS estimate the rate of coincident noise triggers in a combination of detectors labelled by ${\{a\}}$, representing the allowed time window $A$ for coincidences in these detectors and the noise rate estimate $r$ in each detector and each template $i$ at single-detector statistic values $\Breve{\rho}_a$.  The noise models $r_{a,i}(\Breve{\rho}_a)$ are obtained by fitting (detector- and template-dependent) exponential functions to the observed distributions of single-detector triggers. 
The probability densities $p(\vec{\Omega}|S)$ and $p(\vec{\Omega}|N)$ describe the distributions over extrinsic parameters $\vec{\Omega}$ -- the differences in times-of-arrival, phase and the relative amplitude of the recovered triggers -- expected under the signal or noise hypothesis respectively. The last term provides an estimate of network sensitive volume based on the least sensitive detector involved in this coincident event, compared to a constant reference sensitivity across detector combinations. This term then provides a natural weighting in favour of more sensitive detector combinations. Therefore the ranking statistic exploits the information encapsulated in the relative phases, amplitudes, arrival times to the different detectors, noise model and expected distribution of signal \ac{SNR}. 

\subsection{Estimation of statistical significance}
\label{subsec:significance}
The final step in \ac{GW} searches is the determination of the statistical significance of a coincident candidate. PyCBC-based searches do it empirically by first generating background triggers using the method of time-shifts: shift the data from one of the detectors with respect to other detector(s) by a duration greater than the light travel time between them and look for artificial coincidences~\citep{Usman:2015kfa, Davies:2020tsx}. This set of coincidences form the search background, and PyCBC-based searches assign a \ac{FAR} to an apparent event by comparing its ranking statistics with those of the simulated background triggers. The method, by construction, guarantees that the background candidates thus produced can not be entirely due to an astrophysical signal. By repeating this procedure, searches can generate noise background worth more than $10^4$ years with just a few days of data. To minimise background contamination due to signal triggers, the PyCBC toolkit removes triggers from the background within a certain time window around events with large \ac{IFAR}~\footnote{\ac{IFAR} = 1/FAR}.

However, there can be chance coincidences between a signal and a noise trigger. This can significantly contaminate the search background, especially when one of them is loud. The PyCBC toolkit, therefore, hierarchically removes loud triggers to minimise the impact of aforementioned astrophysical contamination while still estimating an unbiased \ac{IFAR}~\citep{TheLIGOScientific:2016pea}.

Currently, PyCBC-based searches analyse data from all possible two and three detector combinations. So it is possible to obtain coincidences of the same event in different detector combinations. The PyCBC pipelines therefore only keep the event with the highest-ranking statistics across detector combinations within a given time-shift window to prevent double-counting. Further, while calculating an event's overall statistical significance, PyCBC searches add the \ac{FAR} from all contributing detector combinations at the ranking statistic of the event, thus accounting for the multiple possible detector combinations.

\section{New Methods for observing intermediate-mass black hole binaries}
\label{sec:new-methods}
At present, two PyCBC-based searches are looking for \ac{GW} signals from \acl{CBC} as recently presented in~\citet{Nitz:2018imz, Nitz:2019hdf, Abbott:2020tfl,  Abbott:2020niy}. The first of the two covers a broad range of parameters ranging from binary neutron stars to \acf{IMBH} binaries, using a hybrid geometric-stochastic template bank~\cite{Roy:2017oul}; we will refer to this search as ``PyCBC-broad''. 
The other search, namely PyCBC-BBH, employs a subset of the templates used in PyCBC-broad and targets specifically binaries having redshifted component masses greater than 5\,\msun\ and mass-ratio $q\in[1,3]$. Thus, both these analyses are not precisely intended to identify \ac{IMBH} binary mergers, and their sensitivities to such signals are correspondingly penalised. Further, both the searches use template banks that are devoid of filter waveforms with a duration less than 150\,ms measured from $\sim20$\,Hz, which further limits their sensitivity towards shorter duration signals~\citep{Salemi:2019ovz, CalderonBustillo:2017skv, Chandra:2020ccy, Abbott:2020tfl}.

This section introduces a new PyCBC-based search that is specifically designed to look for short duration signals arising from putative \ac{IMBH} binaries. Owing to its construction, the search is also sensitive to highly redshifted stellar-mass \ac{BBH} mergers.

\subsection{Search Space and Template Bank}
\label{subsec:template-bank}

The black hole mergers we are mostly targeting form the lower mass range ($10^2-10^3$) of \ac{IMBH} binaries. Signals emitted by these systems are extremely short-lived ($\lesssim 1$s) within the current detector bandwidth, making them difficult to detect. However, since they are a result of \ac{CBC} mergers, they should share morphological resemblances with other observed \ac{GW} signals. Hence like most templated searches, we assume that these signals are adequately described by the dominant harmonic of a quasi-circular, non-precessing binary. This allows us to design and use a template bank that is entirely parameterised by $\vec{\zeta}$.

In what follows, we discuss the salient features of our proposed template bank and how we construct it.
\subsubsection{Bounds on mass-spin space}

\begin{figure}[ht]
  \centering
    \includegraphics[width=\columnwidth]{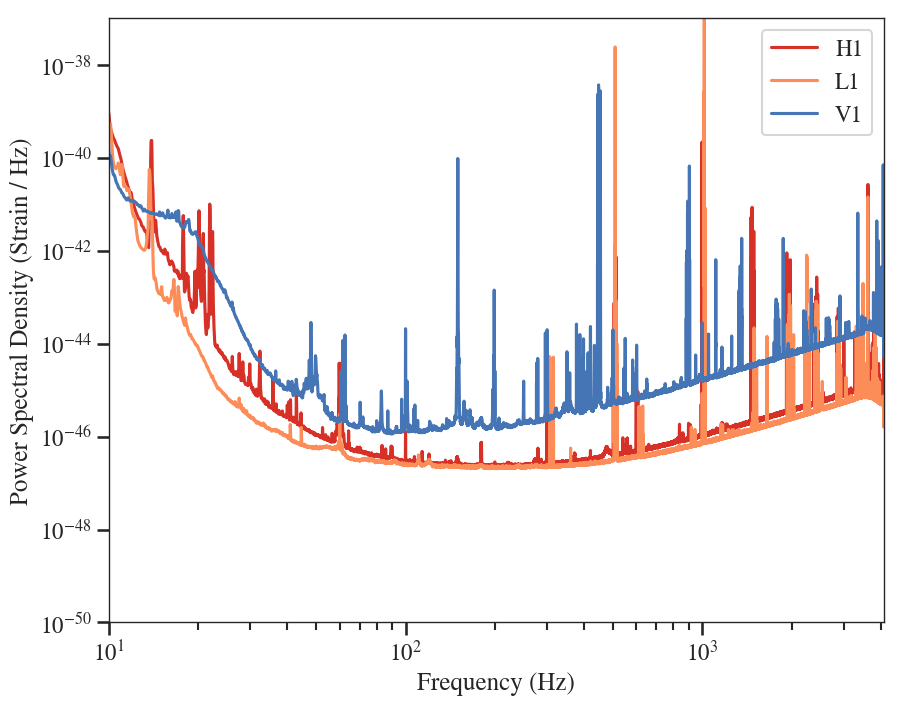}
\caption{\acp{PSD} estimated using data from each of the contributing detectors used in \acl{O3a}. The data is now publicly available at the \href{https://www.gw-openscience.org/data/}{GWOSC} website~\citep{Vallisneri:2014vxa, Abbott:2019ebz}. We use the harmonic mean of these \acp{PSD} to model the detector noise while constructing the bank.}
\label{fig:psd}
\end{figure}

For this work, we construct a bank of filter waveforms imitating \acp{BBH} with detector frame (redshifted) total mass, $M_T(1+z)$, between or close to $100~\msun$ and $600~\msun$. We limit our templates to a mass ratio between $1 \leq q=m_1/m_2 \leq 10$, also imposing that the secondary component has a mass $m_2 > 40\,\msun$. Binaries with more unequal mass ratios range will emit signals with significant higher harmonics, making dominant-mode templates ill-equipped at detecting them~\citep{Harry:2017weg, Mills:2020thr}. Further, systems with detector-frame total mass greater than $600\,\msun$ will emit very few GW cycles within the detector bandwidth; the matched-filter method is only expected to be useful if the number of cycles is significantly $>1$. 
Also, as shown in~\citet{LIGOScientific:2021tfm}, the distance reach to such sources is small because the detector sensitivity degrades rapidly towards lower frequency (see Fig.~\ref{fig:psd}), thus the expected rate of detections is likely to fall off rapidly at higher masses. Further, for such massive systems, the $(2,\pm 2)$ mode is mostly out of the detector bandwidth, implying that templates including only the dominant harmonic won't appropriately recover the signals.  We also conducted preliminary analyses with template waveforms having $M_T(1+z) > 600\,\msun$ and found their noise distributions extended to significantly higher amplitudes and were harder to model or suppress, resulting in a poorer overall search sensitivity.

Since we are ignorant about the spin distribution of the \ac{IMBH} population, we allow $\chi_{1}$ and $\chi_{2}$ of the templates to take on any values between $\pm 0.998$. We don't allow our templates to have misaligned spins~\footnote{Due to spin-spin and spin-orbit coupling, misaligned spins cause orbital precession~\citep{Apostolatos:1994mx}.} as it will violate our underlying assumption of the signal.

Our search domain motivates us to choose the reduced-order representation of spinning effective one-body model, $\texttt{SEOBNRv4}$~\citep{Taracchini_2014, Bohe:2016gbl} as the template bank approximant. Because of the restriction in the waveform model, we bound the dimensionless spin between the specified range and not between theoretical bounds of $\pm 1$. 

\subsubsection{Choice of lower frequency cut-off}
When constructing the bank and calculating the matched-filter \ac{SNR}, we use a lower frequency cut-off of 15\,Hz. We make this choice to balance out a couple of practical considerations. The quadrupolar \ac{GW} frequency at merger scales inversely with the binary's detector frame total mass~\citep{Maggiore:1900zz}. Therefore, particularly for heavier \acp{CBC}, a lower starting frequency for matched filtering equates to better search sensitivity. However, as illustrated in Fig~\ref{fig:psd}, the noise \ac{PSD} of the current advanced detectors increases steeply below 15\,Hz, indicating that a lower starting frequency would not effectively improve the search performance. Furthermore, we observe that lower values of the cut-off frequencies resulted in significantly higher matched-filter \acp{SNR} caused by short-duration glitches~\citep{Soni:2021cjy}. Current signal-noise discriminators, such as $\chi_\textrm{r}^2$ and $\chi_\textrm{sg}^2$, often fail to distinguish these transients from astrophysical population appreciably, especially when the best-matched templates are comparatively short-lived. So, we restrict the matched-filter calculation to the aforementioned lower frequency cut-off to limit the \ac{SNR} of these artefacts. We set the higher frequency cut-off to the maximum frequency of the template both while creating the bank and calculating the matched-filter \ac{SNR}.

\begin{figure}[ht]
  \centering
    \includegraphics[width=\columnwidth]{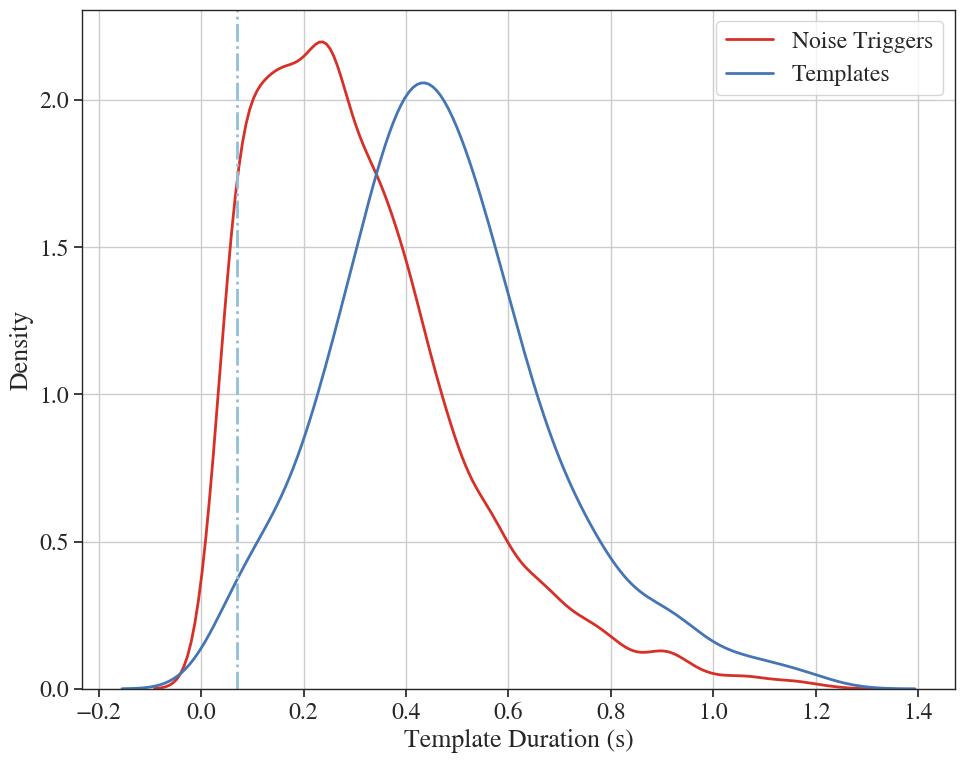}
    \caption{KDE plot showing the distribution of noise triggers [red] with $\rho > 6$ in \ac{O3a}  LIGO-Livingston data while using a template bank without duration threshold. We also show in blue the template duration distribution. The dashed dotted represents the position of the proposed 70ms with respect to the distribution of the noise triggers. The template duration in the negative zone is a KDE artefact.}
\label{fig:glitches}
\end{figure}

\subsubsection{Template duration cut}
The above choices in mass and spin space make our search susceptible to a variety of glitches since short-duration \ac{CBC} templates are well-mimicked by glitches. To demonstrate this, we perform a search over $\sim 32$ days of \ac{O3a} science data using a template bank covering the specified target space and obeying the frequency constraints. As can be seen in Fig.~\ref{fig:glitches} the proportion of the noise triggers increases with a decrease in template duration. After comparing search sensitivity for representative stretches of data using different settings, we empirically excluded any templates having a duration of less than 70\,ms (measured from 15\,Hz) in the detector bandwidth.


\begin{figure}[ht]
  \centering
    \includegraphics[width=\columnwidth]{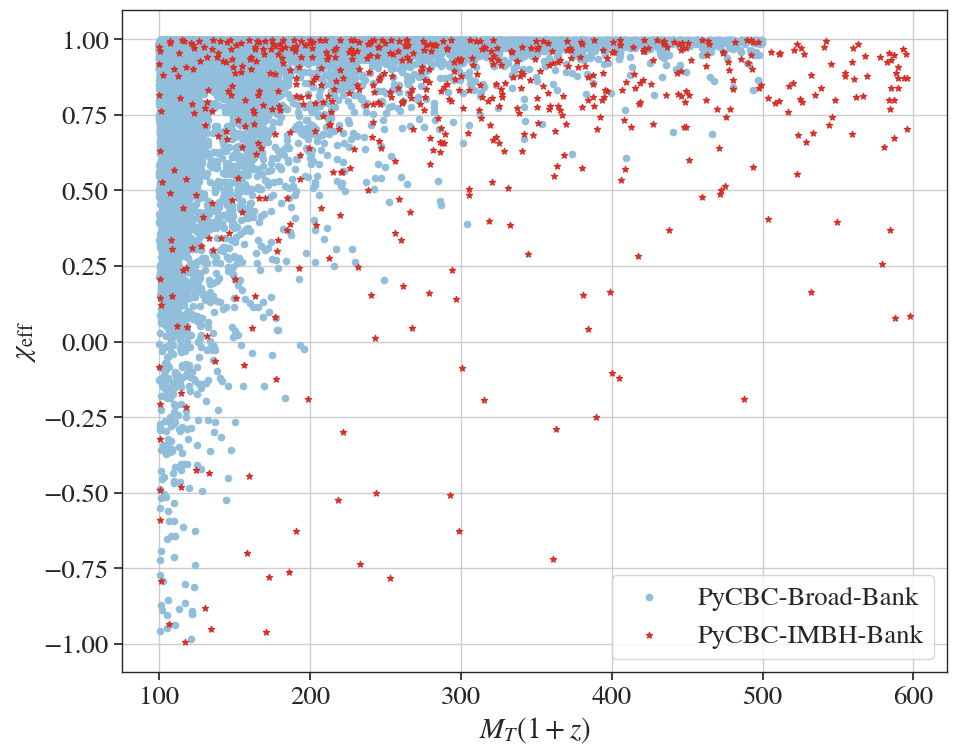}
\caption{The distribution of templates in the redshifted total mass and effective spin space. In red are the templates for the bank used in this work, and in light blue, we represent the subset of templates with redshifted total mass greater than $100\,\msun$, as used in the PyCBC-broad search of~\citet{Abbott:2020niy}. The density of the PyCBC-broad templates in the comparatively lower total mass region is larger due to a higher mass ratio cut of 98 and lower minimum component mass requirements. Despite that, the PyCBC-broad bank has a coarser coverage over shorter-duration templates because of the duration-cut of 150 ms (measured from $\sim 20$ Hz). Note that the PyCBC-broad template bank has no templates beyond $M_T(1+z) = 500\,\msun$ as opposed to PyCBC-IMBH bank.}
\label{fig:bank}
\end{figure}
  
\subsubsection{Stochastic template bank approach}   
Having selected the bounds of our template bank, we now seek to create the bank efficiently. Generally, a template bank can be constructed in two ways --- geometrically~\citep{Owen:1995tm, Owen:1998dk, Babak:2006ty, Cokelaer:2007kx, Brown:2012qf, Harry:2013tca, Canton:2014ena} or stochastically~\citep{Harry:2008yn, Babak:2008rb, Harry:2009ea, Manca:2009xw}, where a hybrid of both methods may be applied to cover a large parameter space~\cite{Roy:2017qgg}.  The geometrical approach uses a quasi-regular lattice and is particularly successful when applied to low-mass systems, for which the merger and ringdown occur outside of the detector bandwidth. Since the post-inspiral phase of the waveform is crucial for our target space, we use the stochastic template placement scheme with a minimal match criterion of 0.99. 

The basic idea of the stochastic template scheme is the following. We begin by randomly selecting a potential template waveforms, $h_{T}$ for our bank $\mathfrak{B}$. Initially, $\mathfrak{B}$ is empty, so we add $h_T$ directly to it. If however there are any templates in  the bank, we evaluate the match, $\mathcal{M}(h_T,h_i)= \max_{\Delta \varphi_c, \Delta t_c} \langle \hat{h}_T | \hat{h}_i \rangle$, between the proposed template and all templates $h_i \in \mathfrak{B}$. Or in other words, we evaluate the effectualness or the fitting factor (FF)~\citep{Apostolatos:1995pj} of $\mathfrak{B}$:
\begin{equation}
    \text{FF} = \max_{h_i \in \mathfrak{B}}~\mathcal{M}(h_T,h_i)~~.
\end{equation}
If the FF of the new template is below the minimal match of 0.99, then the bank $\mathfrak{B}$ is deemed ineffectual, and we add the new template $h_T$ to $\mathfrak{B}$. This generates a new seed bank $\mathfrak{B}'= \mathfrak{B} + h_T $ for the next iteration, and we repeat this process until the rate of rejection of proposed templates reaches a set limit.

\begin{figure}[ht]
  \centering
    \includegraphics[width=\columnwidth]{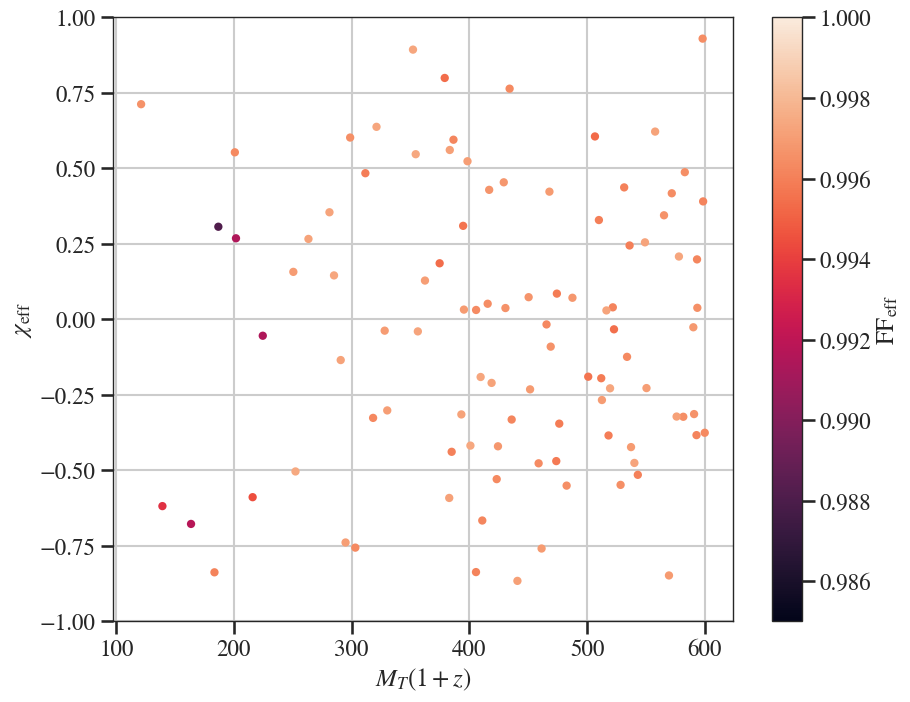}
    \includegraphics[width=\columnwidth]{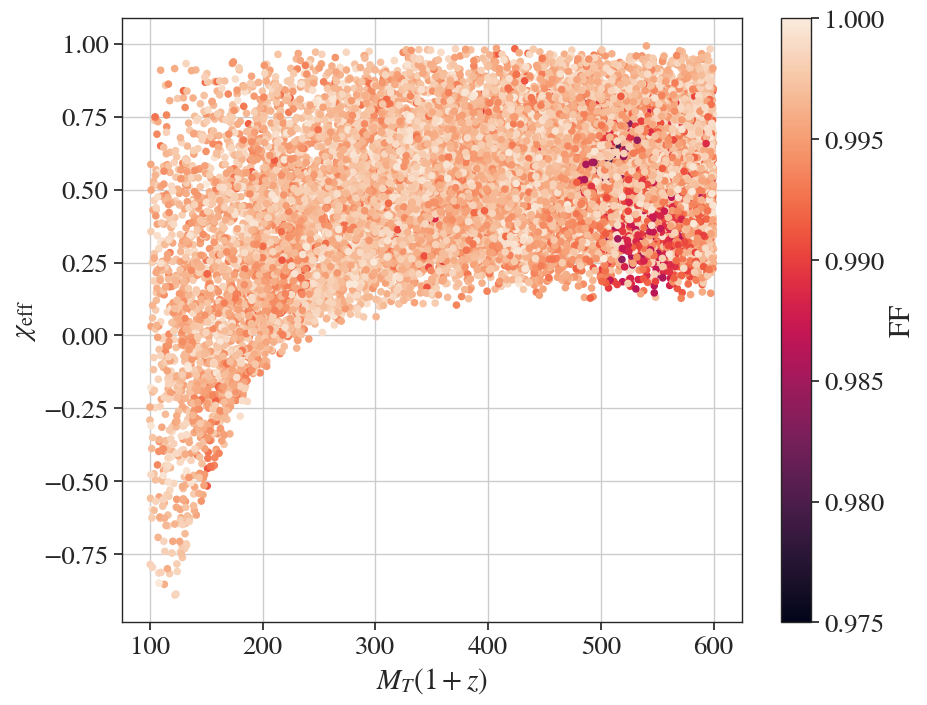}
\caption{Top: The effective fitting factor of 100 simulated signals as a function of total-mass and effective-spin. Note that even though some simulated points lie outside the bank limits, owing to the duration threshold, they still have good $\textrm{FF}_{\textrm{eff}}$. Bottom: Plot showing the fitting factor of 15,600 simulated non-precessing signals lying within the target search space.}
\label{fig:fitting}
\end{figure}

The seemingly high choice of minimal match value ensures an adequate number of templates in the high-mass region of our search space where the signals lack pre-merger power. Fig.~\ref{fig:bank} compares the proposed bank used in this work with that of the one used by the PyCBC-broad in~\citet{Abbott:2020tfl, Abbott:2020niy} in terms of total mass and effective inspiral spin~\citep{Ajith:2009bn, Santamaria:2010yb}~\footnote{The effective inspiral spin $\chi_{\textrm{eff}}=\frac{m_1 \chi_1+ m_2 \chi_2}{m_1 + m_2}$ ranges between $\pm 1$ and gives a measure of the alignedness of the out-of-plane component spin with the direction of orbital angular momentum.}. 

Before performing the search, as a final check, we evaluate the template bank's performance in terms of effective fitting factor~\citep{Buonanno:2002fy}:
\begin{equation}
    \textrm{FF}_{\textrm{eff}} = \sqrt[3]{\frac{\sum_i \textrm{FF}^3(h_i) \rho_{\textrm{opt}}^3(h_i) }{\sum_i \rho_{\textrm{opt}}^3(h_i) }},
\end{equation}
for a hundred different non-precessing simulated signals containing only the dominant harmonic and lying within the search target space. Here, $\rho_{\textrm{opt}}(h_i) = \sqrt{\langle h_i | h_i \rangle} $ defines the \ac{SNR} recovered if the template is an exact replica of the incident signal. As can be seen in Fig~\ref{fig:fitting}, the $\textrm{FF}_{\textrm{eff}} \gtrsim 0.99$ for all the points. Additionally, we also calculate the recovered fitting factor of approximately 15,600 non-precessing simulated quadrupolar harmonic signals lying within our target space. The edge in the bottom panel plot is owing to the duration threshold of 70 ms. Here too, we observe that all of the injections have $\mathrm{FF} \geq 0.975$ signifying that the bank, summarised in Table~\ref{Table: Bank Parameters}, meets the design expectations. 
\begin{table}[ht]
    \centering
    \begin{tabular}{l l}
          Parameter &  \\
         \hline
         $M_T(1+z)$ & $ (100\,\msun,~600\,\msun) $ \\
         $q = m_1/m_2$ & $(1,~10)$ \\
         $m_i$ &  $(40\,\msun,~540\,\msun)$ \\
         $\chi_{\textrm{i,z}}$ & $(-0.998,~0.998)$ \\
         $f_{\textrm{low}}$ & $15$\,Hz \\
         Minimum Match & 0.99 \\
         Minimum Template Duration & 70\,ms \\
         Waveform Model & \texttt{SEOBNRv4\_ROM} \\
         Number of templates & 630 \\
         \hline 
         \end{tabular}
    \caption{Summary of the Template Bank used for the search.}
    \label{Table: Bank Parameters}
\end{table}

We note that applying the duration cut leads to only $\sim 1\%$ fewer templates in the bank but improves the search sensitivity. We demonstrate this in Sec.~\ref{sec:evaluation} when discussing sensitivity computation. Also, as most of the power in the waveforms comes from the merger-ringdown portion of the template, the designed search is sensitive to a wide variety of short-duration signals arising from mergers producing \acp{IMBH}~\citep{CalderonBustillo:2020odh}.

\subsection{Improved Rejection of Instrumental Transients}
\label{subsec:glitch-rejection}
As discussed in Sec.~\ref{sec:review}, an important part of search algorithms for detecting \ac{GW} signals is reducing the number of false alarms caused due to glitches.
\begin{figure}[ht]
  \centering
    \includegraphics[width=\columnwidth]{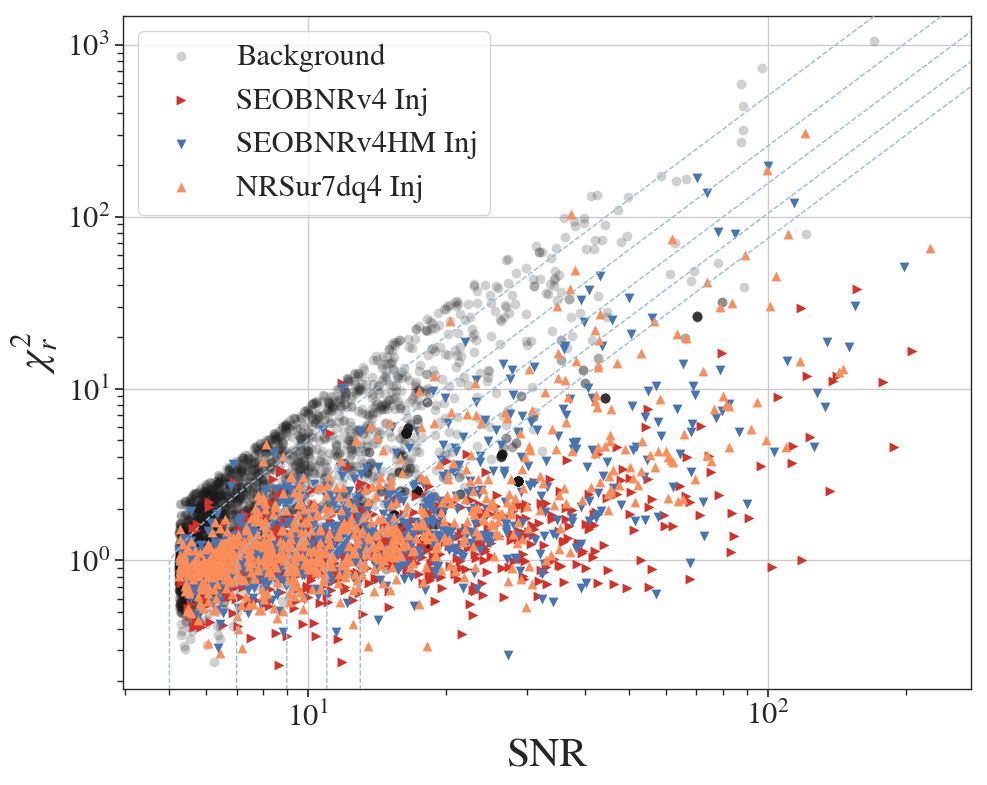}
    \includegraphics[width=\columnwidth]{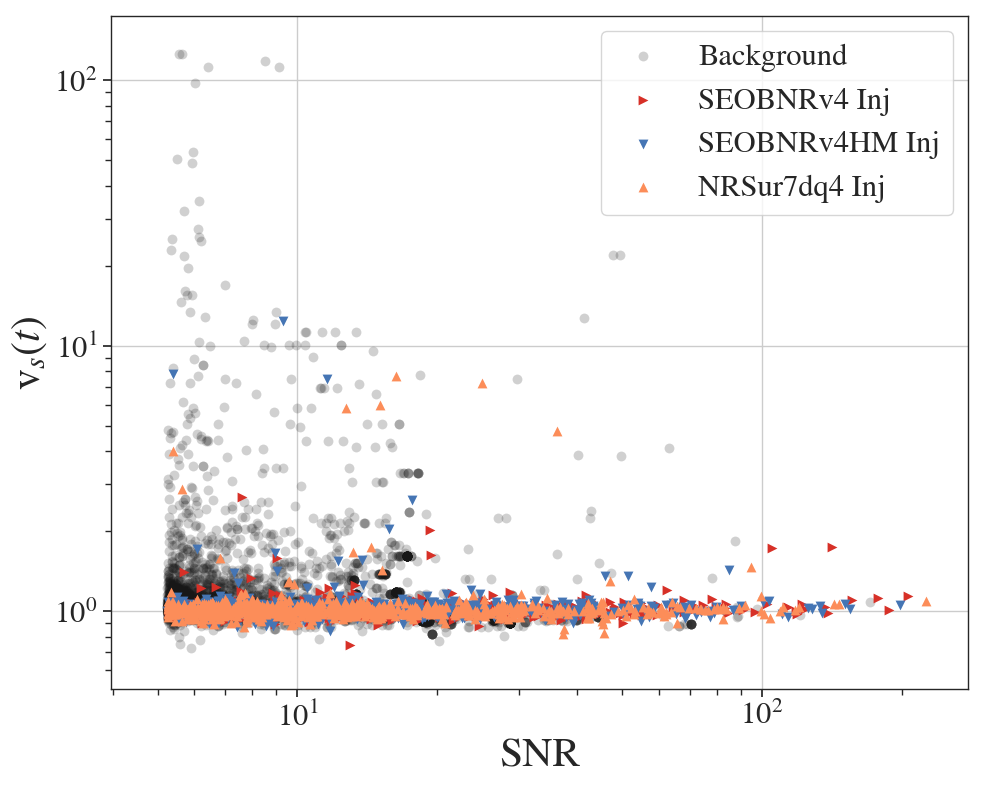}
\caption{Top: Distribution of \ac{SNR} and $\chi_r^2$ of single detector triggers from the representative portion of LIGO-Livingston data from the third observing run. Bottom: Distribution of \ac{SNR} and short-term variation in the \ac{PSD}, v$_s(t)$ from the representative portion of same LIGO-Livingston data from the third observing run. In both the panels, the black dots represent the background triggers which are coincidentally observed in Livingston and Virgo data while the other coloured markers correspond to the recovered injections. The dashed lines in the top panel are contours of $\tilde{\rho} = 5, 7, 9, 11, 13$ respectively.}
\label{fig:vetoes}
\end{figure}
Relative to previous PyCBC searches, our search uses a lower duration threshold and starts filtering from a comparatively low low-frequency cut-off. This makes it susceptible to more glitches, as illustrated in Fig~\ref{fig:glitches}. There is also an absence of data quality flags for Virgo data segments which seriously hampers the search performance/ 
Hence, while pre-processing the data, we choose to use an auto-gating threshold of $50\sigma$ from Gaussian Noise to gate out loud $g(t)$. The threshold used is half of that used in~\citet{Nitz:2018imz,Nitz:2019hdf}. We note that this can also excise out signals of similar loudness, but observed \ac{GW} signals from recent \ac{BBH} mergers have so far not caused such large deviations in whitened detector data.
\begin{figure}[ht]
  \centering
    \includegraphics[width=\columnwidth]{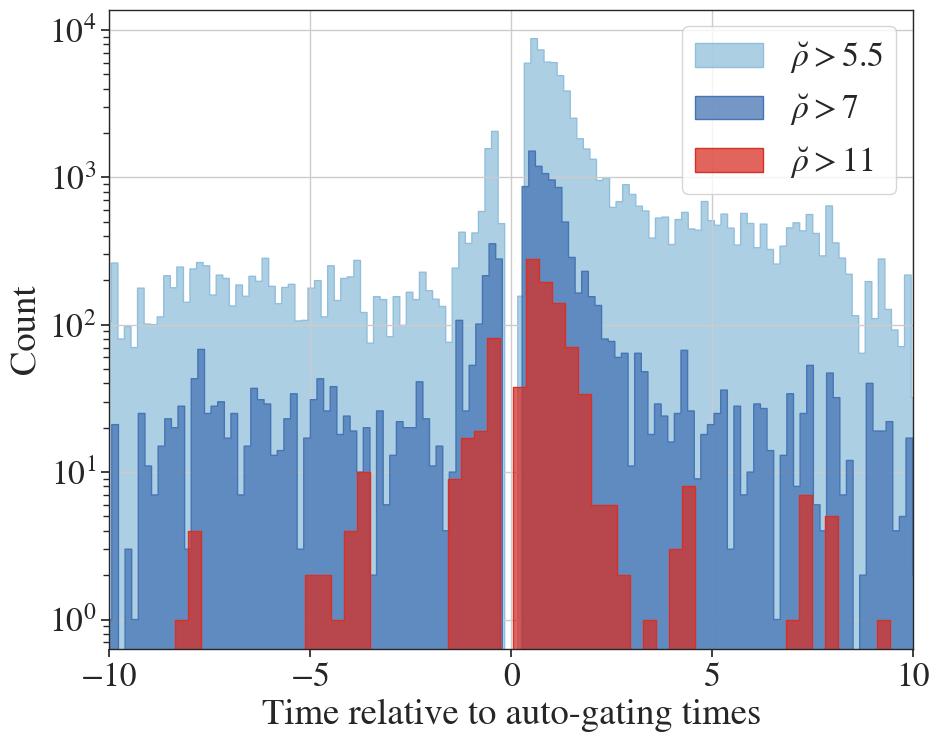}
    \caption{Histogram of triggers close to auto-gated glitches in the Livingston detector. Triggers are removed around the central time within a window of $\pm0.5$\,s around an auto-gate, but we see here that there is a significantly increased trigger rate around auto gates which is not removed.}
\label{fig:gating}
\end{figure}

Even then, while performing the single detector analysis with our proposed bank, we find many loud background triggers that our signal-noise discriminator does not efficiently differentiate. As illustrated in the top panel of Fig.~\ref{fig:vetoes}, these triggers have a large $\tilde{\rho}$ even when having relatively high $\chi_\textrm{r}^2$ and hence are not efficiently suppressed. Further, some background triggers also have large v$_s(t)$, signifying that they are within data segments affected by non-stationary broadband noise. So, we choose to discard any trigger whose $\chi_r^2$ or v$_s(t)$ are greater than 10. The choice is motivated by the separation in the distribution of background triggers and the injection population. We use the \texttt{SEOBNRv4} waveform model to generate quadrupolar quasi-circular non-precessing injections, plotted in red. In blue, we have non-precessing injections containing the full symphony of a signal and they are generated using the reduced-order representation of the \texttt{SEOBNRv4HM} waveform model~\citep{Cotesta:2018fcv, Cotesta:2020qhw}. We generate the remaining orange marked injections with the \texttt{NRSur7dq4}~\citep{Varma:2019csw} waveform model. They contain both the quadrupolar as well as higher-order harmonics of a precessing binary. We provide additional details of these injections in Sec.~\ref{sec:evaluation}.

As can be seen, the above constraint rejects approximately $\sim 3 \%$ of the total injections used. But most of these injections have large single-detector \ac{SNR}($\gtrsim 30$): given current detector sensitivity, we are unlikely to see \ac{IMBH} signals with such loud \acp{SNR}. The distribution of our injections in luminosity distance, and thus in \ac{SNR}, is also not astrophysical, giving relatively more high-SNR signals than in a realistic case, thus we expect the fraction of actual signals rejected to be smaller.

In addition to these veto choices, we also remove triggers from a wider window around auto-gated times of loud non-Gaussianity. During preliminary analysis of \ac{O3a} data, we noticed several instances where the loud glitch prompting the auto-gate was correlated with nearby, lower frequency glitches that had relatively high \acp{SNR} in \ac{IMBH} templates. We can see the effect of this correlation in the histogram of Fig~\ref{fig:gating}, which shows an elevated relative rate of triggers with high $\Breve{\rho}$ in short-duration templates within a few seconds of auto-gated glitches. This effect is mainly visible in both LIGO detectors' data, increasing up to two orders of magnitude, being much subtler for Virgo data. As a first approach to this issue, we veto all triggers between 1 second before and 2.5 seconds after the central auto-gating times of Advanced LIGO data. The tuning limits were chosen empirically and are subject to future improvements. This refinement yields additional suppression of high-statistic background events, leading to an increase in sensitive volume-time, which implies that a \ac{GW} signal belonging to the target space will be detected with an improved significance. We provide details of this study in the next section.
\begin{figure}[thb]
\centering
\includegraphics[width=\columnwidth]{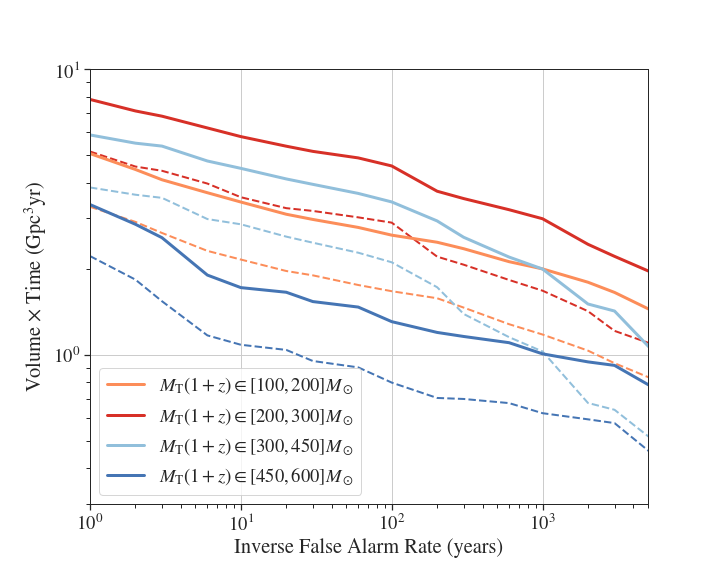}
\caption{Plot showing the sensitive volume-time, VT, that the search can survey before [dashed] and after [solid] applying the duration threshold on filter waveforms. We assess the improvement using a representative population of \acp{GW} from a simulated generically spinning BHs in a binary system injected in $\sim 30$ days of \ac{O3a} data.}
\label{fig:duration}
\hfill
\end{figure}

\begin{figure}[thb]
\centering
\includegraphics[width=\columnwidth]{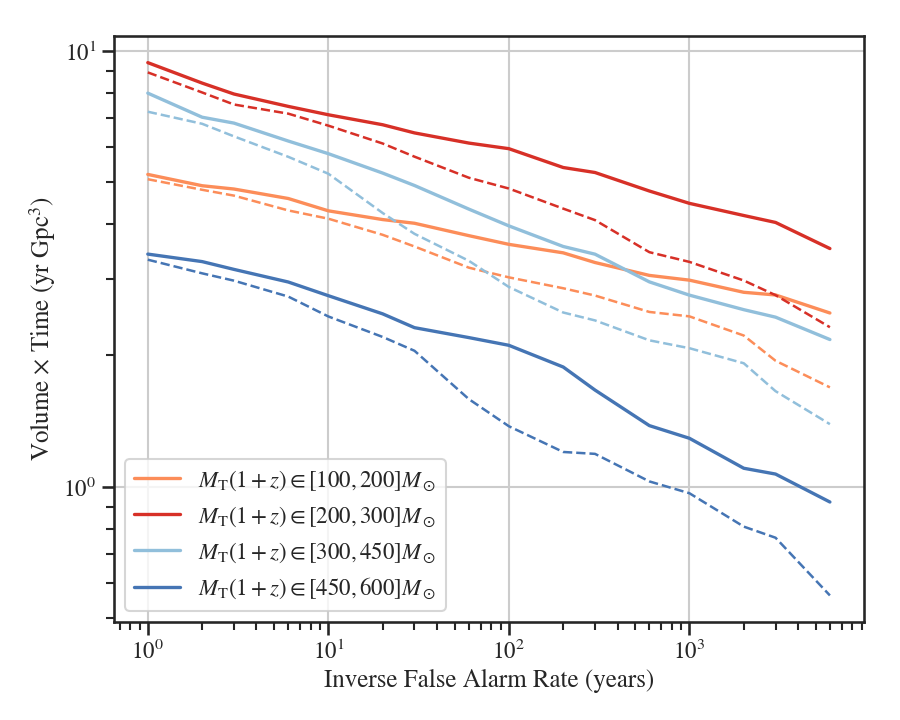}
\caption{Plot showing the sensitive VT before [dashed] and after [solid] the removal of additional triggers around auto-gated times.}
\label{fig:background_with_veto_cut}
\hfill
\end{figure}

\section{Evaluating the Improved Search Technique}
\label{sec:evaluation}

\begin{figure*}[thb]
\centering
\subfigure[Plot showing the ratio of sensitive VT, between PyCBC-IMBH and PyCBC-broad search. ]{\label{fig:imbh_vs_hyper}
\centering
\includegraphics[width=0.75\textwidth]{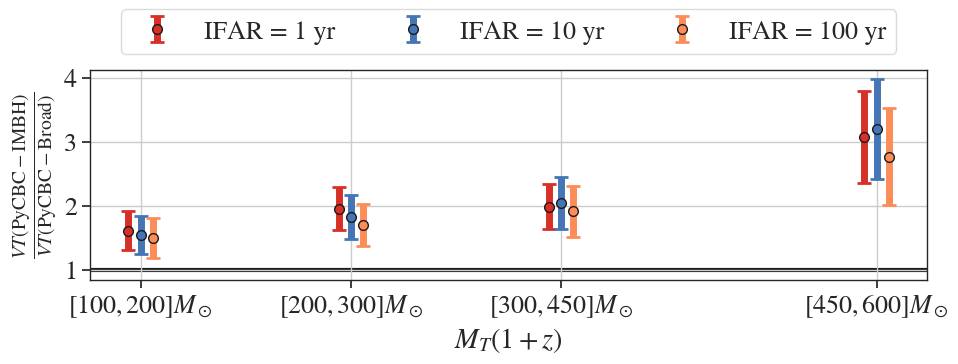}
}
\subfigure[Plot showing the ratio of sensitive VT, between PyCBC-IMBH and PyCBC-BBH search.]{\label{fig:imbh_vs_bbh}
\centering
\includegraphics[width=0.75\textwidth]{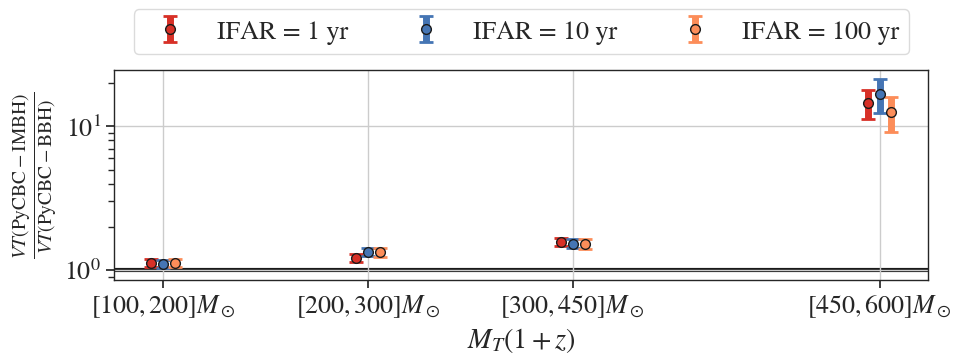}
}
\caption{Plots showing the improvement in sensitivity at different total mass range at different \ac{IFAR}. Both the plots show a significant increment in VT, particularly for signals from relatively heavier binaries.}
\hfill
\end{figure*}

To motivate the various choices made and to quantify the benefit of our restricted analysis as compared to previous PyCBC-based analysis, we resort to an injection campaign: add simulated signals to real or simulated \ac{GW} data as part of the analysis pipeline without any physical actuation and then estimate the sensitivity of the pipeline to the signals added. For our purpose, we inject the simulated signals in $\sim 30$ days of \ac{O3a} data, and we use the sensitive volume-time product, VT, and the sensitive distance as the metrics of the search sensitivity. We estimate these sensitivity metrics using a Monte Carlo method with importance sampling~(See Sec. 4 of~\citet{Usman:2015kfa}).

As mentioned in Sec.~\ref{subsec:glitch-rejection}, we use \texttt{SEOBNRv4}, \texttt{SEOBNRv4HM} and \texttt{NRSur7dq4} as the waveform model for our signals. The simulated binaries have $M_T(1+z)$ in the range $[100,600]\,\msun$ and $q$ between $[1,10]$ except when using precessing injections during which we restrict to $q\leq4$ binaries owing to waveform constraints. The binary spins are isotropically distributed in spin directions as allowed by respective waveform models, and we uniformly distribute the simulated signals in $M_T(1+z)$ and $m_1/(m_1 + m_2)$ space. We also place them uniformly in distances between bounds of 0.5\,Gpc and 11\,Gpc and distribute them isotropically all over the sky and binary orientation parameters. 

Fig~\ref{fig:duration} compares the sensitivity between searches with and without the duration threshold of filter waveforms discussed in Sec.~\ref{subsec:template-bank}. As can be seen, the search using the duration threshold of $70$ ms performs better than the search not using it and the improvement in sensitive VT ranges between a factor of 1.1 to 1.3 at an \ac{IFAR} of 100 years depending on the total mass of the injection. 

The search also benefits from the improved gating setting discussed in Sec.~\ref{subsec:glitch-rejection}. Due to this, we see that the search performance improves significantly, especially for comparatively heavier \acp{BBH} (See Fig.~\ref{fig:background_with_veto_cut}).

Combined, all the proposed optimisations significantly improves the search performance as compared to the already-existing PyCBC-broad search. We illustrate this in Fig.~\ref{fig:imbh_vs_hyper} where one can see a clear increment in sensitive VT between a factor of $1.50\pm 0.31$ and $2.77\pm 0.75$ at an \ac{IFAR} of 100 years depending on the total mass of the system, demonstrating the algorithm's ability to distinguish high-mass signals from noisy transients.

Finally, we also compare the sensitivity of the PyCBC-IMBH search with that of the PyCBC-BBH search used in \citet{Abbott:2020niy} in Fig~\ref{fig:imbh_vs_bbh} for our set of injections. We find that the sensitive VT at an \ac{IFAR} of 100 years varies between a factor of $1.12\pm0.07$ for systems with a total mass between $[100,200]\,\msun$ to $12.61\pm3.52$ for $[450,600]\,\msun$ systems, clearly demonstrating the improvement due to detection optimisation for the \ac{IMBH} binaries. For signals with $M_T(1+z) > 450\,\msun$, the match between the template and signal is quite poor in the PyCBC-BBH search resulting in high $\chi_r^2$ values, which lead to the majority of even high-SNR signals being entirely missed.
\begin{table*}[ht]
  \begin{center}
\begin{tabular}{llllllllll}
Event Name & GPS Time~(s) & $\textrm{IFAR} (\textrm{yr)}$ & \rankingstat{} & $\rho_H$ & $\rho_L$ & $\rho_V$ & $m_1$ (\msun) & $m_2$ (\msun) & $\chieff{}$ \\ \hline
GW$190521\_074359$ & 1242459857.5 & $>$ 43070.23 & 72.55 & 12.00 & 21.14 & - & 60.74 & 40.32 & 0.12 \\
GW$190828\_063405$ & 1251009263.8 & $>$ 14260.16 & 47.92 & 10.04 & 11.20 & - & 60.38 & 40.36 & 0.63 \\
GW$190519\_153544$ & 1242315362.4 & $>$ 9103.08 & 44.02 & 8.37 & 10.51 & - & 100.2 & 62.82 & 0.72 \\
GW$190706\_222641$ & 1246487219.3 & $>$ 8873.62 & 34.81 & 8.73 & 8.36 & - & 149.40 & 44.49 & 0.81 \\
GW$190727\_060333$ & 1248242632.0 & $>$ 8105.10 & 31.58 & 8.37 & 8.11 & - & 60.32 & 40.27 & 0.14 \\
GW$190701\_203306$ & 1246048404.6 & $>$ 5191.03 & 28.64 & 6.01 & 8.76 & 5.85 & 67.68 & 47.02 & -0.48 \\
GW$190915\_235702$ & 1252627040.7 & 2612.10 & 27.87 & 8.41 & 7.50 & - & 73.87 & 49.39 & 0.75 \\
GW$190517\_055101$ & 1242107479.8 & 1153.60 & 26.45 & 6.62 & 7.42 &  - & 58.43 & 48.41 & 0.99 \\
GW$190602\_175927$ & 1243533585.1 &  915.02 & 26.09 & 6.32 & 10.58 & - & 132.67 & 52.10 & 0.47 \\
GW$190521$ & 1242442967.5 &  726.65 & 25.78 & 8.27 & 11.54 & - & 195.46 & 43.06 & 0.50 \\
GW$190503\_185404$ & 1240944862.3 &  397.78 & 27.31 & 9.00 &  7.66 & - & 60.32 & 40.27 & 0.15 \\
GW$190421\_213856$ & 1239917954.3 &  163.39 & 22.66 & 7.88 &  6.25 & - & 58.41 & 53.35 & 0.05 \\
GW$190408\_181802$ & 1238782700.3 &   62.66 & 24.01 & 6.70 &  8.09 & - & 82.52 & 40.07 & 0.98 \\
GW$190513\_205428$ & 1241816086.8 &   19.97 & 20.94 & 8.83 &  7.20 & - & 55.39 & 46.03 & 0.76 \\
GW$190413\_134308$ & 1239198206.7 &    7.10 & 20.32 & 5.46 &  7.50 & - & 79.54 & 56.15 & 0.04 \\
GW$190803\_022701$ & 1248834439.9 &    5.86 & 19.69 & 5.68 &  6.57 & - & 57.33 & 42.78 & -0.08 \\
GW$190929\_ 012149$ & 1253755327.5 & 3.02 & 19.01 & 6.04 & 6.79 & - & 123.03 & 40.67 & 0.15 \\
GW$190413\_052954$ & 1239168612.5 &    1.79 & 18.24 & 5.14 &  6.65 & - & 71.14 & 43.83 & 0.23 \\
GW$190731\_140936$ & 1248617394.6 &    1.56 & 17.09 & 5.61 &  6.00 & - & 58.41 & 53.35 & 0.05 \\
\end{tabular}
    \caption{
    \label{table:events}
Candidate events from the PyCBC-\ac{IMBH} search for compact binary mergers in \ac{O3a} data. Candidates are sorted by \ac{IFAR} evaluated for the entire bank of templates. Finite background statistics limits the \ac{IFAR} estimated for the top six events, and hence we state their lower limits. We list the parameters of the template associated with each candidate. The masses are given in the detector frame while the \ac{SNR} corresponds to the matched-filter \ac{SNR} defined in Eq.~\eqref{eq:SNR}. Note that the listed parameters are not intended as a rigorous estimation of the source parameters, and the reader should refer to \citet{Abbott:2020niy} for the same.}
  \end{center}
\end{table*}

\section{Results}
\label{sec:results}
Using this optimised analysis, we perform a search on the \ac{O3a} data collected during the observation period between 1 April 2019 15:00 UTC and 1 October  2019  15:00  UTC. Table~\ref{table:events} summarises the search results and tabulates them in the descending order of significance in terms of IFAR. We analysed the data separately after sub-dividing them into five independent blocks of duration $\sim 30$ days. The bulk of the events listed are known \acp{BBH} from below the mass range of the template bank, meaning they have been found with reduced significance in templates with around 100\,\msun. We only state the lower limits in \ac{IFAR} of the top six events due to finite background statistics. Apart from the \ac{IFAR} and event id, we also report the GPS time, \rankingstat, the parameters of the best-matched templates and the value of matched-filter \ac{SNR} of the signal in the corresponding detector. Among all the signals reported, only GW$190701\_203306$ is observed as a three-detector event by this dedicated search.

\subsection{Re-detection of GW190521}
\label{subsec:GW190521}
GW190521 was first reported in~\citet{Abbott:2020tfl} where the dedicated cWB-IMBH search recorded the event with \ac{IFAR} 4900 years. However, the offline PyCBC-broad and PyCBC-BBH search failed to detect this signal with enough statistical significance, as mentioned before. Our optimised PyCBC-IMBH search recovers the signal with a network \ac{SNR} of 14.1 and \ac{IFAR} 727 years which is significantly larger than those reported in ~\citet{Abbott:2020tfl} for a PyCBC-based search. This change is due to a culmination of all the modifications that we have made for our analysis. However, the improvement in \ac{IFAR} is not a statistical statement about the search as a whole. The reader should refer to Sec.~\ref{sec:evaluation} for the same.

\section{Conclusion}
\label{sec:conclusion}
We construct and validate a new optimised PyCBC-based search that is capable of detecting both highly redshifted \ac{BBH} mergers and mergers producing low mass range \ac{IMBH} remnants. We test the analysis against previous PyCBC-based searches using a set of generically spinning simulated \acp{BBH} which we inject at different times in the \ac{GW} data collected during the \ac{O3a} run. We find that our analysis can now better separate short-duration \ac{GW} signals from instrumental artefacts because of targeted usage of filter waveforms, optimisation of data quality vetoes and stricter signal-noise discriminators. As a result, the search sensitive VT at an \ac{IFAR} of 100 years improves by a factor of $1.50\pm 0.31$ to $2.77\pm 0.75$ as compared to previous PyCBC-based searches, with the largest increase seen for binaries with total masses between $[450-600]\,\msun$. We also report all the signals with an \ac{IFAR}$> 1.5$ years in our search. However, none of these signals are new, and they have already been reported in \citep{Abbott:2020niy}. 

While we are primarily concerned here with comparing our optimized search for \ac{IMBH} binaries with existing PyCBC searches covering or overlapping with this parameter space, ultimately if several such analyses are performed on the same data, it will be necessary to interpret the joint results consistently.  Using a simplistic approach of selecting only the most significant (lowest \ac{FAR}) candidates from any of the searches, the significance of a given candidate will be diluted by so-called `trials factors', as different searches may yield partly or completely independent sets of noise events, resulting in a reduced sensitivity of the joint search at a fixed \ac{FAR} threshold.  An improved strategy may be obtained by accounting for the relative sensitivities of different searches to the target signal population~\cite{Biswas:2012ty}: in the limit where one search's sensitivity is much larger than others, candidates which are not assigned high significance by that search would be discarded.   More generally, the impact of noise events in a joint search may be mitigated by imposing consistency conditions (derived from studying the recovery of a simulated signal population) on any detection candidates. 

The main limitation of this new search is that it currently relies on waveforms that model the dominant harmonic of a quasi-circular non-precessing binary. A near-term development goal is to make the search capable of detecting \ac{CBC} signals with significant higher harmonic content and develop newer signal-noise discriminators that can better separate a glitch from a short-duration signal. These improvements will enable the search to detect heavier, asymmetric binaries. 

Also, advanced \ac{GW} detectors are expected to observe thousands of \ac{BBH} mergers in the next few years, and the detection rates will go up by several orders of magnitude with the advent of third-generation observatories like the Einstein Telescope~\citep{Maggiore:2019uih, Sathyaprakash:2012jk} and Cosmic Explorer~\citep{Reitze:2019iox} and space-based observatories like LISA~\citep{Sesana:2016ljz}. Their functioning will enable us to detect \ac{GW} signals from even further \ac{BBH} mergers whose detector frame masses can be several orders of magnitude larger than their source mass. Some of these signals may be due to mergers of \acp{IMBH} lying above the pair-instability mass-gap~\citep{Mangiagli:2019sxg, Ezquiaga:2020tns, Tanikawa:2020cca, Mehta:2021fgz}, and detection of such signals will enable us to understand the origins of these binaries better. Also, the detection of mass-gap remnants or components will provide us with a better understanding of the rate at which they are formed. Hence, there is a continuous need to expand the target search space depending on the detector sensitivity to not miss out on potential sources. 

\section{Acknowledgements}
\label{sec:acknowledgements}
The authors would like to thank Shasvath Kapadia, Khun Sang Phukon, Tito Dal Canton and Juan Calderon Bustillo for their detailed comments and useful suggestions. The authors are grateful for the computational resources and data provided by the LIGO Laboratory and supported by National Science Foundation Grants No. PHY-0757058 and No. PHY-0823459. The open data is available in the Gravitational Wave Open Science Center (https://www.gw-openscience.org/ ) a service of LIGO Laboratory, the LIGO Scientific Collaboration and the Virgo Collaboration. The authors also acknowledge the use of the IUCAA LDG cluster, Sarathi, for computational/numerical work. We acknowledge the usage  KC acknowledges the MHRD, Government of India, for the fellowship support. VVO and TD acknowledge financial support from Xunta de Galicia (Centro singular de investigación de Galicia accreditation 2019-2022), by European Union ERDF, and by the ``María de Maeztu'' Units of Excellence program MDM-2016-0692 and the Spanish Research State Agency. AP's research is supported by SERB-Power-fellowship grant SPF/2021/000036, DST, India. GSCD and IWH acknowledge the STFC for funding through grant ST/T000333/1. CM was supported by the STFC through the DISCnet Centre for Doctoral Training.  KS acknowledges the Inter-University Centre of Astronomy and Astrophysics (IUCAA), India, for the fellowship support. This document has LIGO DCC No LIGO-P2100177.

\textit{We want to thank all of the essential workers who put their health at risk during this ongoing COVID-19 pandemic. Without their support, we would not have completed this work. We offer condolences to people who have lost their family members during this pandemic.}

\bibliography{reference}

\end{document}